\documentclass[preprint,12pt]{aastex}

\newcommand{\etal}{{\em et al.}}            

\usepackage{longtable}

\newcommand{\Zsolar}{\mbox{\,$\rm Z_{\odot}$}}
\newcommand{\Msolar}{\mbox{\,$\rm M_{\odot}$}}

\newcommand{\xs}{$\chi^{2}$}
\newcommand{\xsnu}{$\chi^{2}_{\nu}$}

\shorttitle{LMXBs in NGC 5102}
\shortauthors{Kraft \etal}

\begin{document}

\title{A Chandra Observation of the Nearby Lenticular Galaxy NGC 5102: Where are the X-ray Binaries?}
\author{R. P. Kraft}
\affil{Harvard/Smithsonian Center for Astrophysics, 60 Garden St., MS-67, Cambridge, MA 02138}
\author{L. A. Nolan, T. J. Ponman}
\affil{School of Physics and Astronomy, University of Birmingham, Birmingham, B15 2TT, UK}
\author{C. Jones}
\affil{Harvard/Smithsonian Center for Astrophysics, 60 Garden St., Cambridge, MA 02138}
\author{S. Raychaudhury}
\affil{School of Physics and Astronomy, University of Birmingham, Birmingham, B15 2TT, UK}

\begin{abstract}

We present results from a 34 ks Chandra/ACIS-S
observation of the LMXB population and the hot ISM
in the nearby (d=3.1 Mpc) lenticular galaxy NGC 5102,
previously shown to have an unusually low X-ray luminosity.  
We detect eleven X-ray point sources within the 
the $D_{25}$ optical boundary of
the galaxy (93\% of the light),
one third to one half of which are likely to be background AGN.
One of the X-ray sources is coincident with the optical nucleus
and may be a low-luminosity AGN.
Only two sources with an X-ray luminosity greater than 10$^{37}$ ergs s$^{-1}$ in
the 0.5-5.0 keV band were detected, one
of which is statistically likely to be a background AGN.
We expected to detect 7 or 5 such luminous sources if the XRB population scales linearly with the
B band or J band magnitudes, respectively, of the host galaxy.  
By this measure, NGC 5102 has an unusually low number of XRBs.
The deficit of LMXBs is even more striking because some of these sources
may in fact be HMXBs.  NGC 5102 is unusually blue for its morphological type,
and has undergone at least two recent bursts of star formation only $\sim$1.5$\times$10$^7$
and $\sim$3$\times$10$^8$ years ago.
We present the results of optical/UV spectral synthesis analysis and
demonstrate that a significant fraction ($>$50\%) of the stars in this galaxy
are comparatively young ($<$3$\times$10$^9$ years old).  We discuss the
relationship between the XRB population, the globular cluster population,
and the relative youth of the majority of stars in this galaxy.
If the lack of X-ray binaries is related to the relative youth of most
of the stars, this would support models of LMXB formation and evolution that require
wide binaries to shed angular momentum on a timescale of Gyrs.
We have also analyzed archival HST images of NGC 5102, and
find that it has an unusually low specific frequency
of globular clusters ($S_N\sim$0.4).  The lack of LMXBs could also be explained
by the small number of GCs.
We have also detected diffuse X-ray emission in the central $\sim$1 kpc
of the galaxy with an X-ray luminosity of 4.1$\times$10$^{37}$ ergs s$^{-1}$
in the 0.1-2.0 keV band.  This hot gas is most likely a superbubble created by
multiple supernovae of massive stars born during the most recent star burst, and
is driving the shock into the ISM which was inferred from
previous [O III] $\lambda$5007 and H$_\alpha$ observations.

\end{abstract}


\section{Introduction}

Early-type galaxies have been known to be luminous X-ray sources
since the Einstein era.  The X-ray emission from the most
massive elliptical galaxies is dominated by
emission from thermal coronae.  These coronae may be primordial
and enriched with metals from outflows created by supernovae or stellar winds,
or they may be entirely built up from such outflows, with the primordial
gas being blown off during a large, early burst of star formation \citep{dav90,dav91,ewan01}.
There is a well known correlation between the X-ray luminosity
of an elliptical galaxy and its total mass (quantified
by its blue magnitude), although there is more than
a factor of 100 dispersion around the mean.
For less massive galaxies, and for massive
galaxies that have lost much of their corona through tidal
or ram pressure stripping,
the contribution of the population of X-ray binaries (XRB)
to the total X-ray luminosity becomes increasingly important.  The contribution of the
XRB population to the X-ray spectra of early galaxies
was noted in Einstein and ASCA observations \citep{kim92,mat97}.
The Chandra X-ray Observatory (CXO) has observed a large number
of elliptical and S0 galaxies and has resolved the brightest XRBs from the hot ISM
(e.g.~\citet{sar00,kra01,kun02} among many others).
The XRB populations of galaxies with a range of masses and
morphological types have now been studied in detail, and
it is clear that the variance in the number of XRBs per
unit optical luminosity in early galaxies is larger than that
expected based on counting statistics alone.
This variance is likely related to the origin and evolution of
the XRBs, subjects on which there is still considerable debate.

There is general agreement that the creation of XRBs must be a continuous, ongoing
phenomenon in the life cycle of early galaxies
because the lifetime of all XRBs is considerably less ($\sim$10$^9$ yrs for
low mass X-ray binaries (LMXBs) and $\sim$10$^7$ yrs for high mass X-ray
binaries) than the age of their host galaxy \citep{gri02}.
The specific mechanism for XRB formation is still a subject of contention.
Some have argued that most or all of the XRBs in elliptical galaxies
originate in globular clusters because they are the only regions where
the stellar density is sufficient for neutron stars to be captured
in N-body interactions.  Recent Chandra observations of NGC 4472, NGC 1399,
and other massive early galaxies
have shown that a significant (40-60\%) fraction of the XRBs are coincident
with globular clusters, supporting this idea.  In this scenario,
the variance in the number
of XRBs can then be explained as differences in the globular cluster
populations.  However, the ellipticals for which a large XRB/GC correlation 
have been demonstrated are located in rich environments and have exceptionally 
large globular cluster specific frequencies (GCSFs).
Even in early-type galaxies with unusually large
GCSFs, at least half of the XRB population is 
not directly accounted for by this GC/XRB correlation. 
Others have argued that the XRB population is related
to the bulk of the stars in the galaxy, and that the variance in the
XRB populations is related to the star formation history of the host
galaxy \citep{whi98,wu01}.  The fact that the
majority of the LMXBs in the galactic center
region of the Milky Way and the M31 bulge do not appear to be associated with GCs
supports this argument.  Recent analysis of the X-ray point source population from a sample of early galaxies
suggests both mechanisms are important \citep{irw04}.

In this paper, we present results from a 34 ks Chandra/ACIS-S observation of
the S0 galaxy NGC 5102.  NGC 5102 is a small
($M_B$=-17.45) nearby (d=3.1$\pm$0.15 \citep{mcm95})
lenticular galaxy and is a member of the Centaurus A group \citep{isr98}.
The luminosity of this galaxy, and therefore presumably the mass,
is similar to the LMC.  
NGC 5102 has been previously observed by Einstein and ROSAT \citep{for85,irw98,ewan01},
where it was noted that its integrated X-ray luminosity is considerably less
than one would expect from a simple extrapolation of its mass compared with
the more luminous ellipticals.
The bulge of this galaxy is unusually blue for its morphological
type, and it is believed to have
undergone one or more recent bursts of star formation (within
$\sim$10$^8$ years ago), although
spectral synthesis modeling suggests that most or all of the stars are
relatively young ($<$5$\times$10$^9$ yrs) \citep{pri79,roc87,deh97}.
Note that unlike most other early-type galaxies,
it is possible that NGC 5102 has a significant HMXB population because of
the recent star formation.  In order to better understand the link between the
XRB populations, the GC population, and the star formation history of the
stellar population, we have analyzed archival HST data to search for GCs and
made a spectral stellar synthesis study of the optical/UV spectra of the galaxy
to constrain the star formation history.

The purpose of this observation was two-fold.  The primary goal was
to resolve the XRB population from the hot gas
and quantify the nature of the X-ray
deficit in this unusually blue galaxy.
Because of the proximity of the galaxy and the sensitivity
of the CXO, we have been able to detect point sources below $L_X$$\sim$10$^{36}$ ergs s$^{-1}$,
much deeper than most Chandra observations of more massive early-type
galaxies at the distance of the Virgo cluster.
We wanted to compare the XRB population of this galaxy with other early-type
galaxies observed with Chandra.
Naively, one would expect that the recent bursts of star formation might
enhance the XRB population relative to older ellipticals due to the presence
of the additional population of HMXBs. 
On the contrary, there are surprisingly few XRBs.
A secondary goal of this observation was to resolve any hot gas that may
be present in the galaxy from the XRB population, and to determine its
thermodynamic state.

This paper is organized as follows.  Section 2 contains a brief
discussion of the observation and data preparation.  The X-ray
point source (XPS) population and its luminosity function
are presented in section 3.
Spectral and spatial analysis of the hot ISM is described in section 4.
A discussion of the significance of the lack of XRBs, particularly
as it relates both to the stellar ages and star formation history as
derived from optical/UV spectral synthesis and to the GC population, 
is presented in section 5.  We conclude with a brief discussion
and implications of our results in section 6.
We assume a distance to NGC 5102 of 3.1 Mpc throughout this paper \citep{mcm95},
and J2000 coordinates are used in all figures and tables.

\section{Observation Log}

\begin{figure}
\plotone{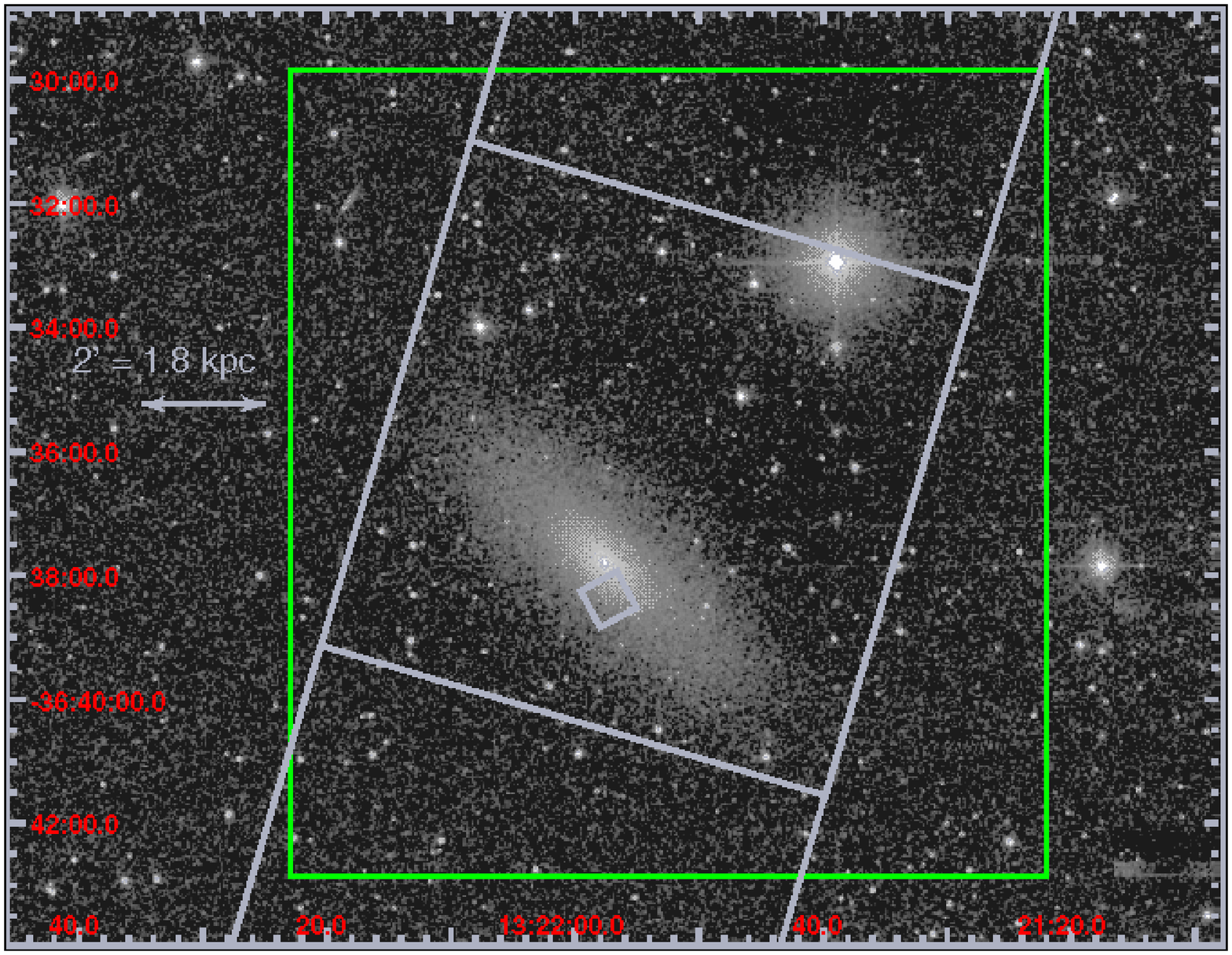}
\caption{The Chandra/ACIS-S FOV overlaid onto a 2MASS J band
image of NGC 5102.  North is up, east is to the left.  The
coordinates in this and all other figures are J2000.
The diamond in the center of the image is
the position of best focus on the S3 chip,  The chips to the NNE and SSW are
the S4 and S2 chips, respectively.  The green box denotes the boundary of
the region searched for point sources.}\label{fovovl}
\end{figure}

The lenticular galaxy NGC 5102 was observed by the Chandra X-ray
Observatory for 34217 $s$ with the
ACIS-S detector in very faint (VF) mode.
A summary of the optical properties of NGC 5102 and the Chandra observing log
are contained in Table~\ref{n5102tab}.
Very faint mode event filtering was applied to the data
to reduce the background to the lowest possible levels.
Hot columns, events near node boundaries, short 
term transients, and the streaks on the S4
chip were also removed.  The background above
2 keV was examined for periods of flaring and none
were found.  The level of the background in the 2-10
keV band agrees well with the nominal blank-sky ACIS-S background.
A 2MASS J-band image with the FOV of the ACIS S2, S3, and
S4 CCDs overlaid is shown in Figure~\ref{fovovl}.

\section{X-ray Point Sources}

Individual point sources were detected with the CIAO
program {\em wavdetect} in the 0.5-5.0 keV band.
The detection sensitivity was chosen so that there is approximately
one false detection for every million pixels.
The source list generated by {\em wavdetect} was visually compared
with a raw image to confirm that every source it detected was
indeed a source.  We likewise inspected the raw image to ensure that all sources
visible by eye were detected by the program.
We detected 55 sources within the intersection
of a 12.2$'$$\times$13.0$'$ box centered on the galaxy and the
ACIS-S FOV (see Figure~\ref{fovovl}).
Based on the sensitivity, we expect one of these to be a false source.
To convert the observed count rate to flux, we have assumed a 5 keV
thermal bremsstrahlung spectrum with galactic ($N_H$=5$\times$10$^{20}$ cm$^{-2}$)
absorption, a spectrum typical of LMXBs.
For this spectrum, a count rate of 10$^{-3}$ cts s$^{-1}$ in the
0.5-5 keV band corresponds to a flux (unabsorbed) of 7.04$\times$10$^{-15}$ ergs cm$^{-2}$ s$^{-1}$ in the
same band.  We define the boundary of the galaxy to be the
$D_{25}$ isophote \citep{deV91}.  The major and minor effective diameters 
at this isophote are 8.7$'$ and 2.8$'$, respectively, and the apparent diameter of the
effective aperture is 0.79$'$.  Assuming that the light is distributed
according to the de Vaucouleurs $R^{1/4}$ law, this ellipse contains approximately 93\% of the light.
The positions, count rates and uncertainties, and X-ray luminosities (if the source is located within
the $D_{25}$ boundary) of the sources are
shown in Table~\ref{bigtab}.  The $D_{25}$ isophote is contained entirely within the
S3 chip.  Several of the sources outside this boundary
appear to be coincident with foreground stars in 2MASS and DSS images.
We note that several of the sources (labeled with a dagger in
Table~\ref{bigtab}) were detected on the S2 and S4 chips.
For these sources (assuming the same spectrum and correcting for
vignetting), a count rate of 10$^{-3}$ cts s$^{-1}$ in the
0.5-5 keV band corresponds to a flux (unabsorbed) 
of 1.06$\times$10$^{-14}$ ergs cm$^{-2}$ s$^{-1}$ in the
same band. 

\begin{figure}
\plotone{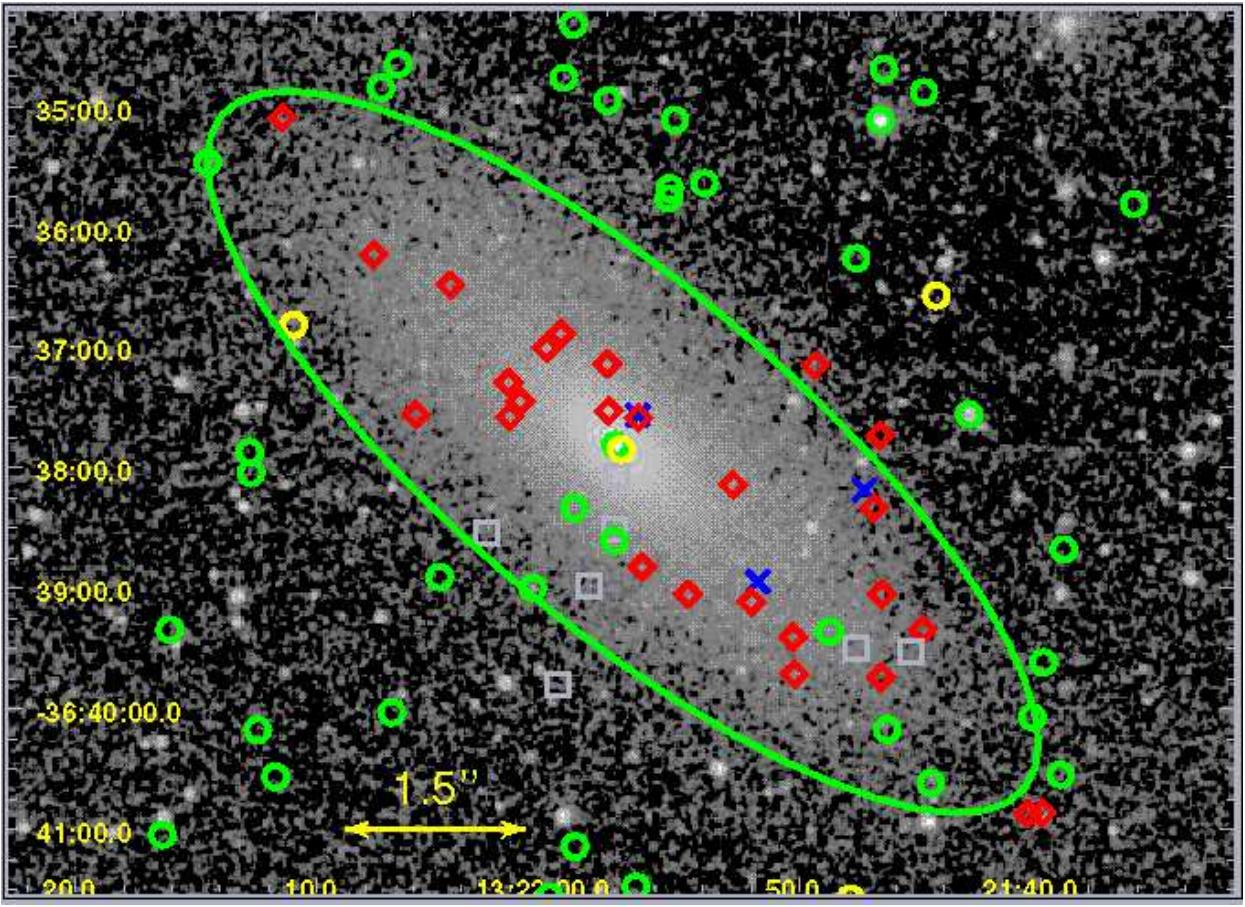}
\caption{Positions of the X-ray points sources overlaid onto
a 2MASS J band image of NGC 5102.  The large green ellipse denotes
the optical boundary ($D_{25}$) of the galaxy, the yellow and green circles the
XPSs with $L_X\geq 10^{37}$ ergs s$^{-1}$ and $L_X<10^{37}$ ergs s$^{-1}$, respectively,
the red diamonds are planetary nebulae, and the white boxes are HII
regions (the later two taken from \citet{mcm95}).  The blue Xs represent
GC candidates.}\label{psovl}
\end{figure}

Only eleven X-ray point sources (XPSs) are contained within the $D_{25}$ boundary
of the galaxy (all on the S3 chip),
and only two sources with $L_X\ge$10$^{37}$ ergs s$^{-1}$ in the 0.5-5.0
keV band.  A 2MASS J band image with the positions of all of the sources overplotted
is shown in Figure~\ref{psovl}.
The positions of HII regions and planetary nebulae (PN) have also
been overplotted for visual comparison \citep{mcm95}.  There are no co-incidences
between the XPSs and any of the HII regions or PN.
Two of the XPSs are located near the optical center of the
galaxy.  One of the sources is one of the two
sources with $L_X>$10$^{37}$ ergs s$^{-1}$, and the other, less luminous source
is coincident with the optical center/nucleus.  It is
possible that this second source is a low-luminosity AGN (LLAGN),
although no other evidence of nuclear activity has been reported in the literature.
There is a group of sources approximately 3.5$'$ north of the
center of the galaxy, but whether this group is related to NGC 5102
is not known.

\begin{figure}
\plotone{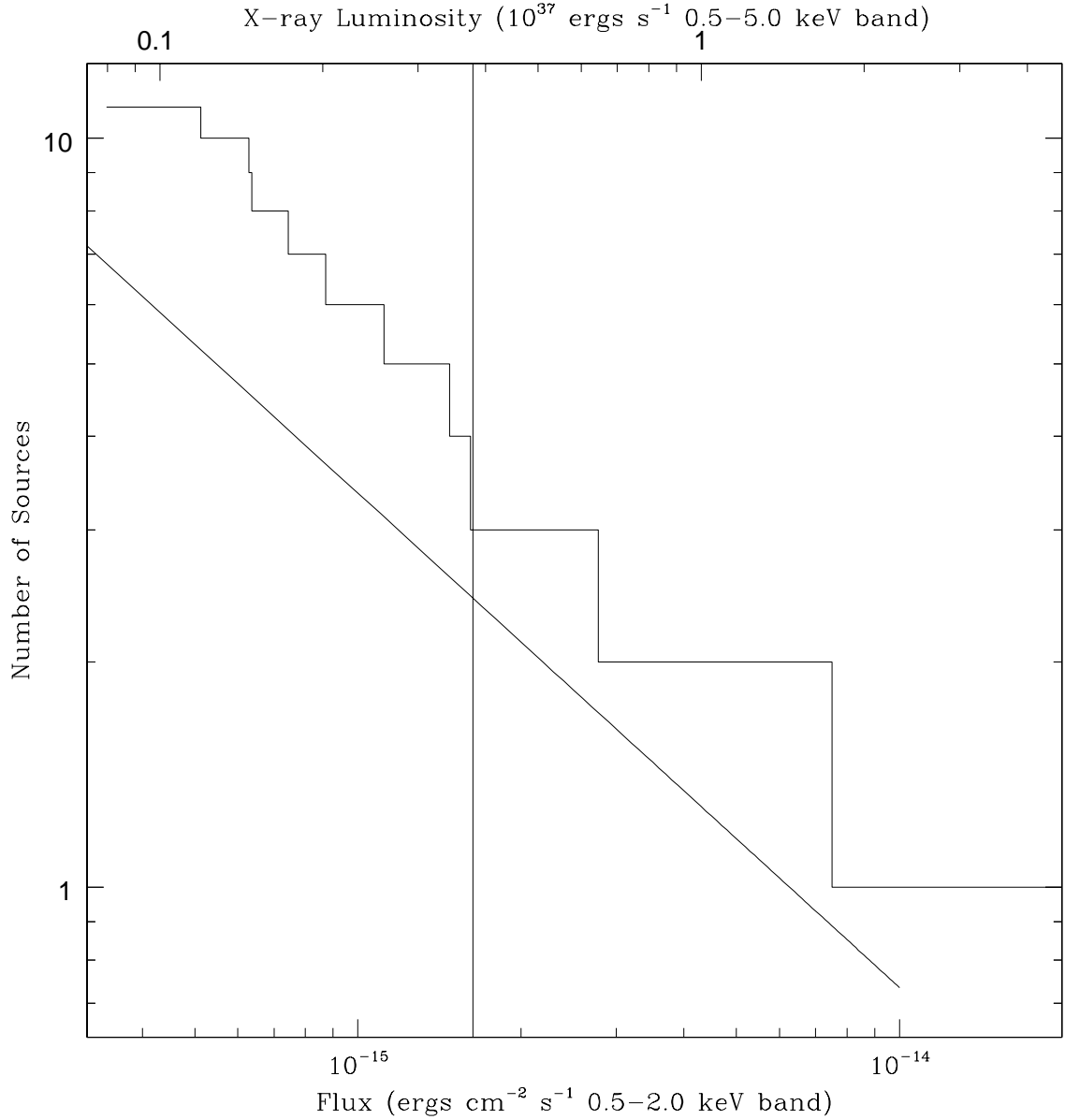}
\caption{Cumulative luminosity function (the histogram) of X-ray point
sources detected in elliptical region (within the $D_{25}$ isophote) described in text.
The log(N)-log(S) of the background AGN is plotted
as the continuous curve \citep{toz01}.  The corresponding X-ray luminosity of the
sources in the 0.5-5.0 keV band (assuming the source to be at a distance of
3.1 Mpc) is shown across the top.  The solid vertical line denotes the flux
corresponding to 16 counts in our observation.  Above this line, we are complete
and unbiased (see text for detailed discussion).}\label{xpslf}
\end{figure}

The luminosity function (LF) of the XPSs contained within the
optical boundary of the galaxy is plotted
in Figure~\ref{xpslf}.  A significant fraction of these are likely
to be background AGN unrelated to NGC 5102.  We have overplotted an estimate of the number of
background AGN onto Figure~\ref{xpslf} based on number counts from deep observations of the
Chandra Deep Field South \citep{toz01}.  Note that the source fluxes have been scaled 
in this figure to the 0.5-2.0 keV band in order to simplify comparison 
with the deep survey results.  
Our point source detection is complete and unbiased, defined as a 4$\sigma$ measurement of the
luminosity (see \citet{kra01} for details)
above a luminosity of $L_X$=4$\times$10$^{36}$ ergs s$^{-1}$. 
Above this luminosity, two of our three sources are statistically likely to be
background AGN.
Half of the sources in the LF shown in
Figure~\ref{xpslf} were detected with less than 10 counts.  A detailed statistical
comparison between the LF of the galactic sources and the background
AGN would require an extensive Monte Carlo simulation to assess the
effects of various biases introduced by low number counting statistics
(see \citet{ken03} for a detailed discussion of these biases and references).  
The details of this are not
significant for our discussion below, and such an analysis is beyond
the scope of this paper.  We conclude that of the 11 detected sources, 
it is likely that a significant fraction are background objects unrelated to NGC 5102.
Based on an examination of the HST images (described below), source \#17 is likely
to be an unrelated background galaxy.
Beyond the optical boundary of the galaxy, the surface density of
X-ray sources is approximately consistent with the deep survey
results.  The integrated X-ray luminosity of the eleven sources contained within
the optical boundary of the galaxy is 5.6$\times$10$^{37}$ ergs s$^{-1}$
in the 0.5-5 keV band.

NGC 5102 has considerably fewer XRBs than expected based on a linear extrapolation
from more massive early-type galaxies.  To quantitatively demonstrate
this we compare the number of XRBs with $L_X\geq $10$^{37}$ ergs s$^{-1}$ to
that observed in Cen A \citep{kra01,kra05} and estimated from an extrapolation of the
XRB luminosity function in NGC 4472 \citep{mac03}.
As shown in Figure~\ref{xpslf}, we have detected two sources within the boundary of
the galaxy with $L_X\geq$10$^{37}$ ergs s$^{-1}$, one of which is statistically likely to be
an unrelated background AGN.  It is possible that both or neither of these
sources are related to NGC 5102, but for the purposes of this analysis, we will
assume that one of the sources is related to the galaxy, and one is an unrelated background
object.  
The choice of $L_X\geq$10$^{37}$ ergs s$^{-1}$ as the limiting luminosity for our comparison
is well above our completeness threshold so that the biases described in the
previous paragraph are insignificant.  An XRB with a luminosity of
10$^{37}$ ergs s$^{-1}$ at the distance of NGC 5102
corresponds to a source with 43 counts in our $\sim$34 ks
observation.  We have not missed any sources 
with $L_X>$10$^{37}$ ergs s$^{-1}$ associated with the galaxy, and the uncertainty in the estimation
of luminosity is an insignificant source of systematic error.

The apparent and absolute luminosities in the J and B bands
of NGC 5102, Cen A, and NGC 4472 are summarized in Table~\ref{galsum}.
Approximately 110 XRBs were detected in Cen A with $L_X>$10$^{37}$ ergs s$^{-1}$
(this is probably a slight underestimate because the distant halo
was beyond the FOV of these observations), and
we estimate that there are $\sim$220 such XRBs in NGC 4472 based on a
extrapolation of the luminosity function (Figure 4 of \citet{kun02})
to this luminosity.  Scaling these
values by the ratio of B band luminosity, we expect to detect 6 such LMXBs in
NGC 5102.  We have detected one.  As stated above, NGC 5102 is unusually blue because
of bursts of recent star formation.  A similar scaling by J band luminosity predicts 5 such
LMXBs in NGC 5102.  The Poisson probability
of observing 1 XRB if the mean is 5 is not unreasonably low ($\sim$3.4\%),
but improbable.  We are certainly in the regime of small number statistics, but
emphasize that this result is in general agreement with that of \citet{irw98} who
found that the X-ray luminosity (above 0.5 keV) of NGC 5102 is less than
predicted based on extrapolation from higher mass objects.  
Possible explanations for this result are discussed below.
The lack of XRBs cannot be a general result related to lower mass elliptical
and/or lenticular galaxies because there are several similar galaxies that have a hard
X-ray component in their spectra that are consistent with an extrapolation
to higher mass galaxies \citep{ewan01}.

The XLF below 10$^{37}$ ergs s$^{-1}$ gives us additional
information, but it is more difficult to make a quantitative comparison
because there are few observations
of other galaxies to this limiting luminosity.
We have detected approximately 6-9 objects in NGC 5102
with $36<\log (L_X ({\rm ergs\ s}^{-1}))<37$ (see
the discussion above).
If we extrapolate the XLF of Cen A down to a limiting
luminosity of 10$^{36}$ ergs s$^{-1}$ and scale by blue
luminosity, we would expect about fifty sources with $36<\log (L_X ({\rm ergs\ s}^{-1}))<37$ in
NGC 5102.  We note that this is only an extrapolation, and that
the power law index in this luminosity range
is not known, although there is a suggestion that it flattens out
(see Figure~8 of \citet{kra01}).  CXO observations of the bulge of
M31 detect a clear flattening of the power law index \citep{kon03}
in this luminosity range.
The paucity of sources in NGC 5102 in the luminosity range
$36<\log (L_X ({\rm ergs\ s}^{-1}))<37$ may or may not be
typical of early galaxies in general, but there is certainly not a
large ($\sim$25-50) population of sources in this luminosity range, with a drastic
steepening of the LF power law index above 10$^{37}$ ergs s$^{-1}$.

\section{Diffuse Emission}

\begin{figure}
\plotone{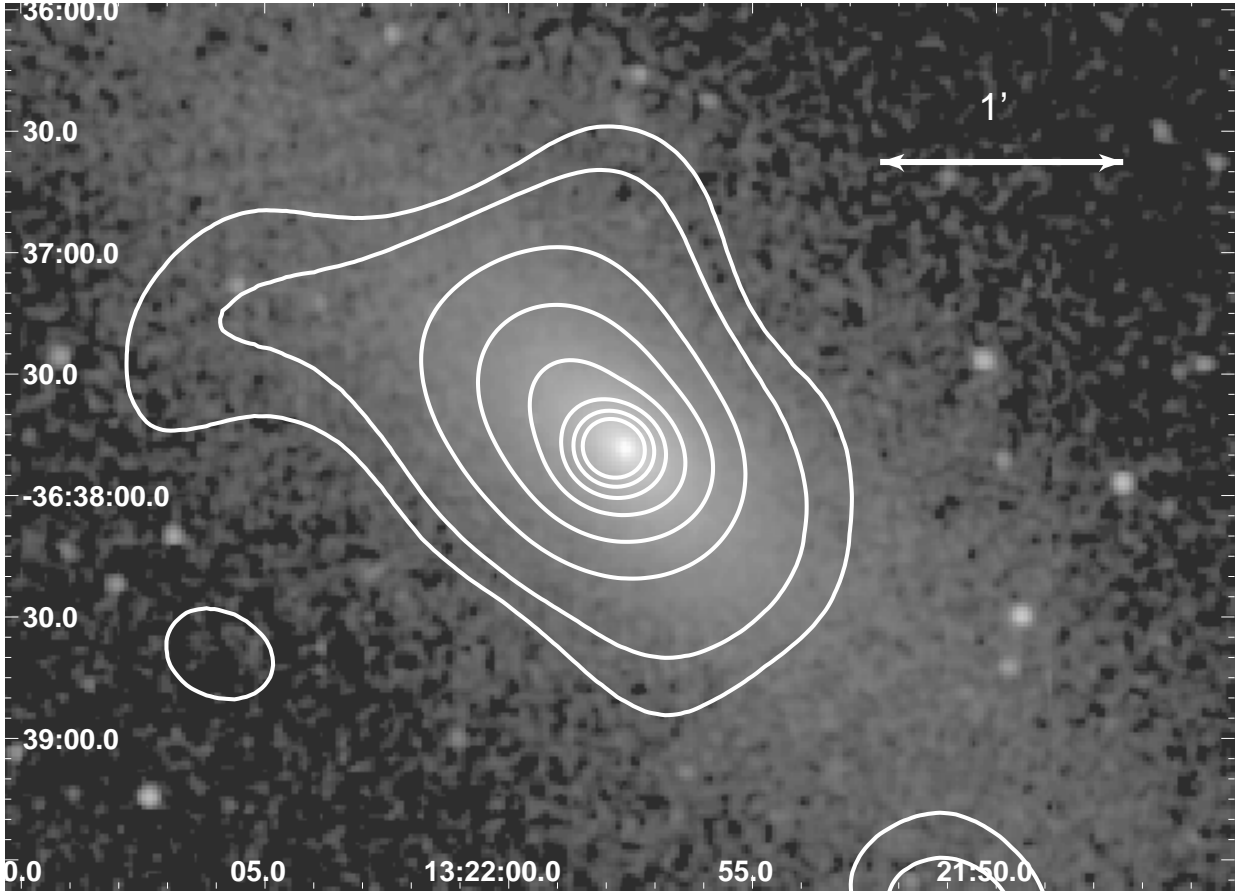}
\caption{Contours from adaptively smoothed, exposure corrected,
background subtracted X-ray image of NGC 5102 in the 0.5-2.0
keV band overlaid onto a 2MASS J band image.  The contours correspond 
to X-ray count rates of 1.0, 1.5, 2.8, 4.9, 7.9, 11.7, 16.4, and
22.0$\times$10$^{-7}$ cts arcsec$^{-2}$ s$^{-1}$ above background.}\label{ismovl}
\end{figure}

We have created an adaptively smoothed, exposure corrected X-ray image in the
0.5-2.0 keV band with the point sources removed.  Contours from this
images are overlaid onto a 2MASS J band image in Figure~\ref{ismovl}.
The diffuse emission is centrally peaked and centered on the optical/IR
center of the galaxy.  There are too few counts in the diffuse emission
to constrain model parameters of a surface brightness profile fit (e.g.
a $\beta$-model) in a meaningful way.  
We extracted a spectrum of the unresolved emission from a region
$1'$ (900 pc) in radius centered on the optical center of the galaxy.  Background
was determined from an identical sized region on the S3 chip away from
the galaxy.  Hypothesising that the emission is due to hot gas,
we fit the spectrum with an absorbed APEC model using XSPEC (V11)
holding the absorption constant at the galactic value (5$\times$10$^{20}$ cm$^{-2}$).
The elemental abundance was also held fixed at 0.4 relative to solar.  Additional fits were
performed with one or both of these parameters free, but they were not constrained in
a meaningful way, although they generally preferred a low
($<0.05$) value for the abundance.
The low abundance is commonly interpreted as indicating the presence of
multi-temperature gas with spatially varying abundances \citep{fab03}.

The best fit temperature is $0.31^{+0.23}_{-0.07}$ keV (90\% confidence for
abundance fixed at 0.4 times solar), and the
density of the gas is $n_H$=7.2$\times$10$^{-3}$ cm$^{-3}$ assuming the gas is uniform 
throughout this region and that $n_H$=1.18$n_e$, appropriate for an ionized plasma with
subsolar abundances.
From Figure~\ref{ismovl}, it is clear that the surface brightness
of the gas is centrally peaked.
The X-ray luminosity of the gas is 4.1$\times$10$^{37}$ ergs s$^{-1}$ (unabsorbed) in
the 0.1 to 2.0 keV band.
We note that this is more than an order of magnitude less than the luminosity
estimate of \citet{irw98}.  Much of this difference can be accounted for
by differences in the assumed distance to NGC 5102
and the measured gas temperature.
It is possible that there is additional cooler gas present in NGC 5102 to which
the ACIS-S instrument is not particularly sensitive.
The sensitivity of the ACIS-S has been gradually reduced because of the
build-up of contamination.  
In any case, the ROSAT PSPC had considerably more effective area
than Chandra/ACIS-S below 0.5 keV.  The addition of a second,
lower temperature ($T<0.1$ keV) component marginally improves the quality of the fits,
but the limited statistics prevent us from making a quantitative statement.
The low temperature argues against this diffuse emission actually being a
population of unresolved, lower luminosity X-ray binaries.

This hot ISM may be related to the ring-like feature seen in optical emission lines
by \citet{ber76} and \citet{mcm95}.  They detected an asymmetric ring-like structure
in [O III] $\lambda$5007 and H$_\alpha$ emission extending approximately 0.7 kpc
from the nucleus.  The preferred model of \citet{mcm95} for the origin of this
emission was shock excitation from an expanding super bubble.  The expansion
velocity was estimated at 60 km s$^{-1}$ from optical line ratios.
We suggest that the X-ray gas is this superbubble which was created
as a result of the most recent epoch of star formation.  The expansion of
this hot gas is responsible for shock heating
the ISM and the appearance of the emission line structures.
The total mass and thermal energy of the hot gas is $\sim$10$^6$ $M_\odot$ and
$\sim$10$^{54}$ ergs, respectively.  These values are close
to the estimates of the total mass and energy of the starburst in the
nuclear region \citep{mcm95}.
The fact that the X-ray emission is centered on the nucleus supports the
hypothesis that it is related to a nuclear starburst, and not to some type of
phenomenon in the bulge.  This bulge phenomenon was invoked by \citet{mcm95} to explain how
the ring-like structure could have survived 2$\times$10$^8$ yrs against differential
rotation.  At a velocity of only 60 km s$^{-1}$, the structures could only be
about 10$^7$ yrs old.  More recent optical and UV observations suggest that the
time since the last starburst was only 15 Myrs \citep{deh97}.

\section{LMXB Formation: Star Formation History versus Globular Clusters}

The apparent lack of XRBs in NGC 5102 gives information
about the details of their formation and evolution.
The standard paradigm is that the majority of the luminous XRBs in elliptical and lenticular galaxies
are low mass X-ray binaries:  neutron stars accreting matter via Roche lobe overflow from
a nearby late type stellar companion, such as Sco X-1 or Cyg X-2 in our galaxy.
The details of the origin and evolution of these X-ray binaries are still poorly understood
and the subject of considerable debate.  
The lifetime of a neutron star accreting matter from a late-type dwarf with
$L_X>$10$^{37}$ ergs s$^{-1}$ is only $\sim$10$^9$ yrs, so that either the XRB
phenomenon is intrinsically transient (i.e. we are presently observing only a small number
of the XRBs present), XRBs are being continuously created during the lifetime
of a galaxy,  or the emission of many of the XRBs
is beamed so that we have greatly overestimated their luminosities
and therefore underestimated their lifetimes.  This third
phenomenon does not appear to be important for the vast majority
of luminous LMXBs in our galaxy
and it will not be considered further here.
The temporal variability and evolution of a large population of XRBs outside the
Milky Way is only now beginning to be studied with Chandra.  
It is possible that we have been unlucky and observed NGC 5102 during a period
when an unusually large number of LMXBs are in quiescence.
However, unless the temporal properties of the LMXBs in NGC 5102 are for some reason
different from those in other early galaxies, this is improbable.
Long term monitoring the XRB population of the Milky Way and Cen A suggests that any
snapshot of the entire XRB population is a representative sample even though
many of the individual sources may be highly variable \citep{gri04, kra05}.
Thus, differences in the XRB populations among galaxies is likely related to
their formation and evolution.

The formation and evolution of LMXBs is thought to proceed along one of two general lines.
One the one hand, it has been proposed that the majority of XRBs
in early galaxies are formed in globular clusters, the only region
where the stellar densities are sufficient to create such systems
via N-body interactions \citep{whi02,mac03}.
Recent observations of nearby, massive elliptical galaxies support this claim
and find that 30-60\% of the luminous XRBs are coincident with
globular clusters, although only 3-4\% of the GCs contain X-ray
sources \citep{ang01,mac03}.
\citet{whi02} found a positive correlation between $L_{LMXB}/L_{OPT}$ ratios
and globular cluster specific frequency.
As an alternative, it has been argued that the LMXB population of early galaxies
is related to the star formation history of the host galaxy \citep{whi98,wu01}.
A variety of scenarios have been developed to explain the evolution
of close binary systems and the subsequent creation
of LMXBs \citep{web83,kal96,kal98}.
Generally, the LMXB begins as a wide binary with a large
mass ratio.  The more massive star rapidly proceeds through its
lifecycle and explodes as a supernova.  The orbit of the binary
decays, perhaps through magnetic braking, until Roche lobe overflow
occurs and the LMXB is born.
The luminosity function contains information on
the birth/death rate of XRBs and on past epochs of star formation.
The XRBs should be distributed roughly as the optical starlight in this
scenario.  This hypothesis is supported by the fact that the majority of the LMXBs
in the bulges of both the Milky Way and M31 do not appear to be related
to GCs.  Both components are likely to play an important role,
the relative importance of each for any individual
galaxy may depend on a wide range of factors such as environment, star formation
history, etc.  

In order to assess the applicability of these two general models 
to the unusual case of NGC 5102,
we have analyzed archival HST imaging data to study the GC
population, and have used stellar spectral synthesis analysis on the
UV/optical spectrum of the galaxy to constrain it's star formation history.
We find that the majority of the stellar population is young, and
that NGC 5102 has an unusually small number of GCs.  Either of these
could explain the deficit of X-ray sources.

\subsection{Stellar Spectral Synthesis}

Given the unusual color of this galaxy, it is interesting
to determine its star formation history and ask
whether the deficit of XRBs is related to this history.
In order to evaluate any potential relationship between the
deficit of XRBs in NGC 5102 and its star formation history,
we have estimated the age of the stellar populations
of the galaxy by fitting a
two component stellar population evolutionary synthesis model
to the UV-optical spectrum of this galaxy.
In a future publication, we will present the results of a more
comprehensive comparison of the XRB population and the star
formation history derived from spectral synthesis of a large sample of early
galaxies.  The UV and optical data of NGC 5102 were taken from several archival sources as described
in Table~\ref{datatable}.  Adjacent spectral
sections were normalized to unity in the region of
wavelength overlap spliced together. The data are
corrected for Galactic reddening. The errors in the flux are the
observed errors for the IUE data ($\lambda\le$3100 \AA), and are
estimated at 6$\%$ of the observed flux for the optical data ($\lambda>$3100 \AA).
There is sufficient data in the long-baseline spectrum
to allow the robust disentanglement of multiple stellar populations.
A two-component stellar population evolutionary synthesis model was
constructed, using the instantaneous starburst models of Jimenez et
al. (1998). These have ages ranging from 0.01 $-$ 14 Gyr, and
metallicities 0.01, 0.2, 0.5, 1.0, 1.5, 2.5 and 5.0 \Zsolar. Age,
metallicity and fractional contribution by (stellar) mass of each
component of the composite model were allowed to vary as free
parameters.   The details of this technique are presented elsewhere
(Dunlop \etal, 2005, in preparation).

\begin{figure}
\includegraphics[width=4.5in,angle=-90]{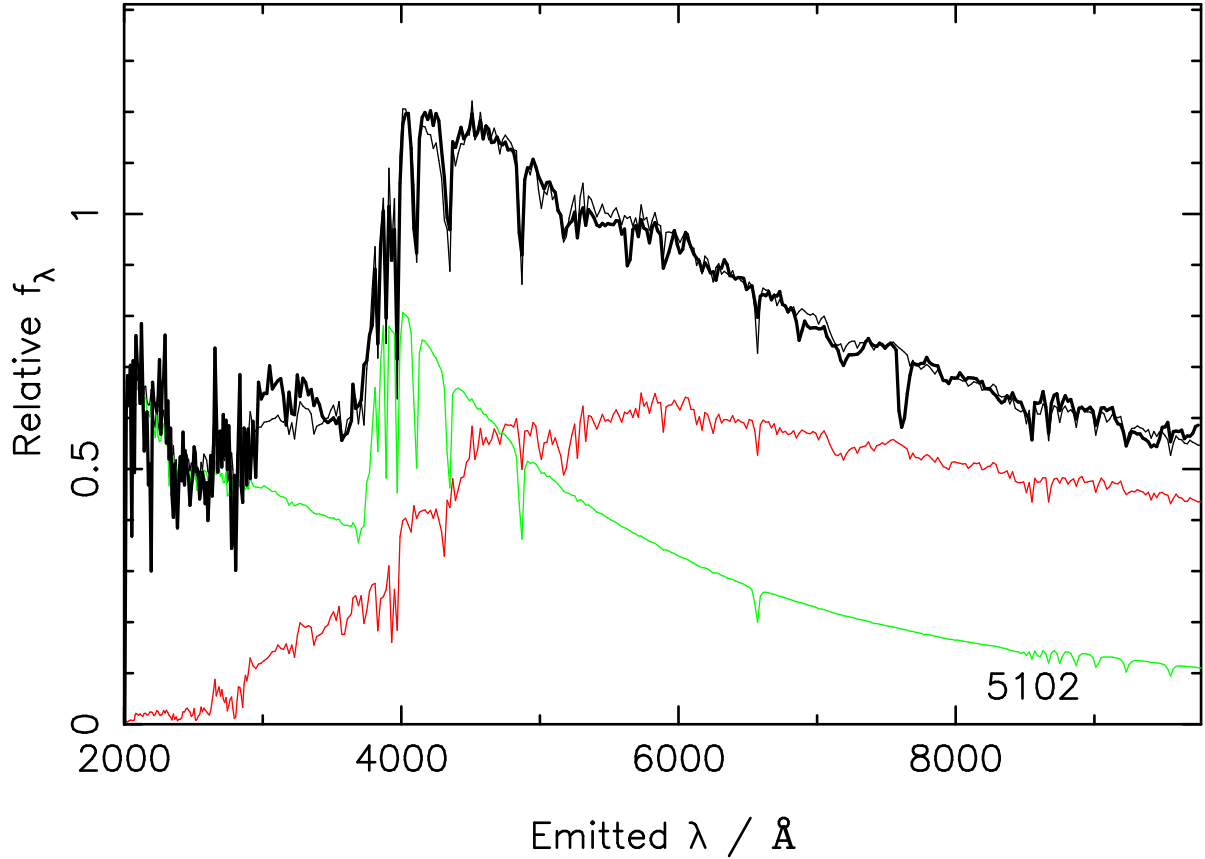}
\caption{The best-fitting two-component model (thin black line) superimposed
over the spectrum of NGC 5102 (thick black line). The two component populations
are also shown; the dominant population (3 Gyr, Z = 1.5 \Zsolar) is in red, and the lesser population
(0.3 Gyr, Z = 0.2 \Zsolar) is in green.}\label{specfig}
\end{figure}

\begin{figure}
\includegraphics[width=4.5in,angle=-90]{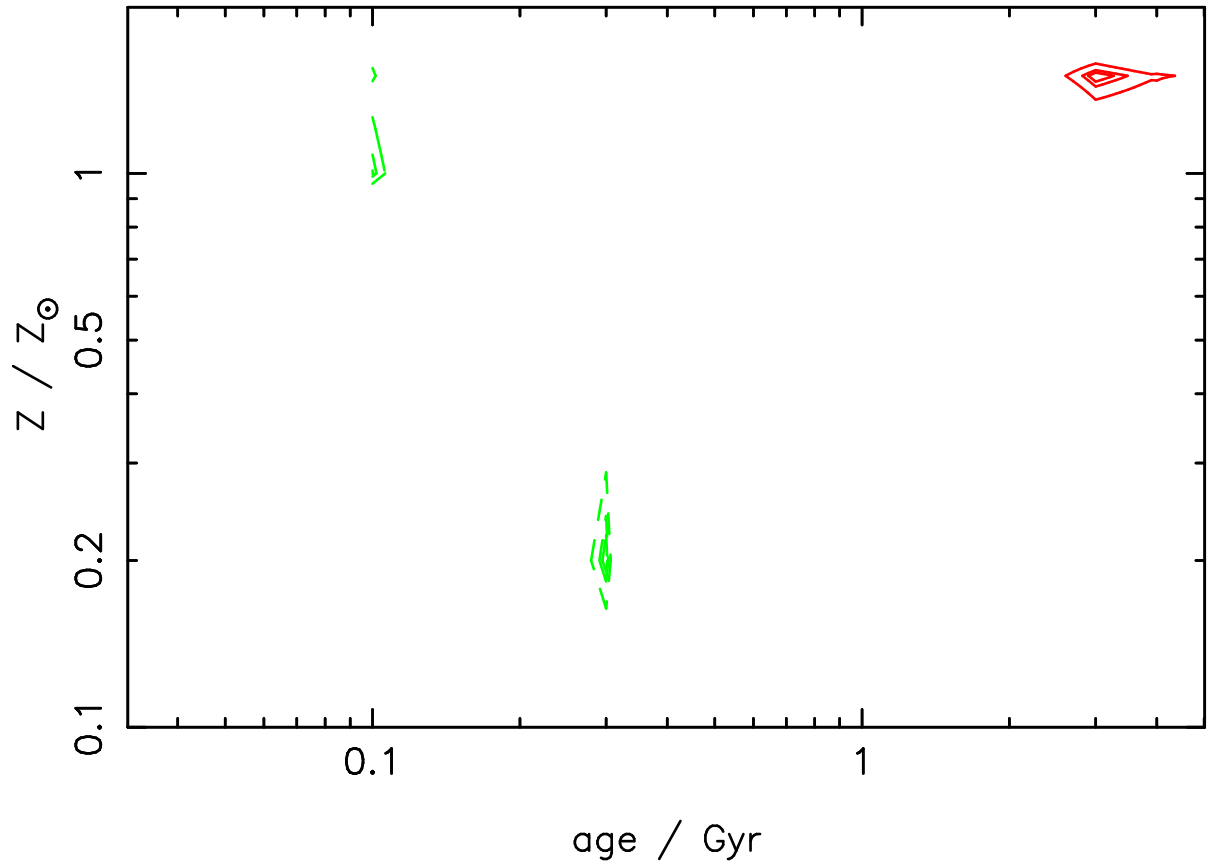}
\caption{Contour plot of constant relative likelihood. The contours contain
68.3\%, 90\% and 95.4\% relative likelihood. The dominant population contours
are in red, and the secondary population contours are green.}\label{specfig2}
\end{figure}

Figure~\ref{specfig} shows the results of our long-baseline spectral
fitting. Using this technique, we have been able to disentangle
two major stellar populations in NGC 5102, and to robustly constrain
their ages, metallicities and relative stellar masses, as listed in
Table~\ref{spectable}. The dominant (M/M$_{gal} = 97\%$) population
has an intermediate age (3 Gyr), and super-solar metallicity (1.5
\Zsolar). Other authors have estimated the age of the younger
population, which resides in the nucleus, and these are consistent
with the 0.3 Gyr, 0.2 \Zsolar\ secondary population that our
two-component model fitting finds. The uncertainty contours of the
parameters is shown in Figure~\ref{specfig2}.  Results from \citet{pri79,roc87,bic88}
constrain the age to within $\sim 0.1 - 0.5$ Gyr. \citet{bic88} also determines that
this population has a sub-solar metallicity, Z $=$ 0.3 \Zsolar.

\begin{figure}
\includegraphics[width=4.5in,angle=-90]{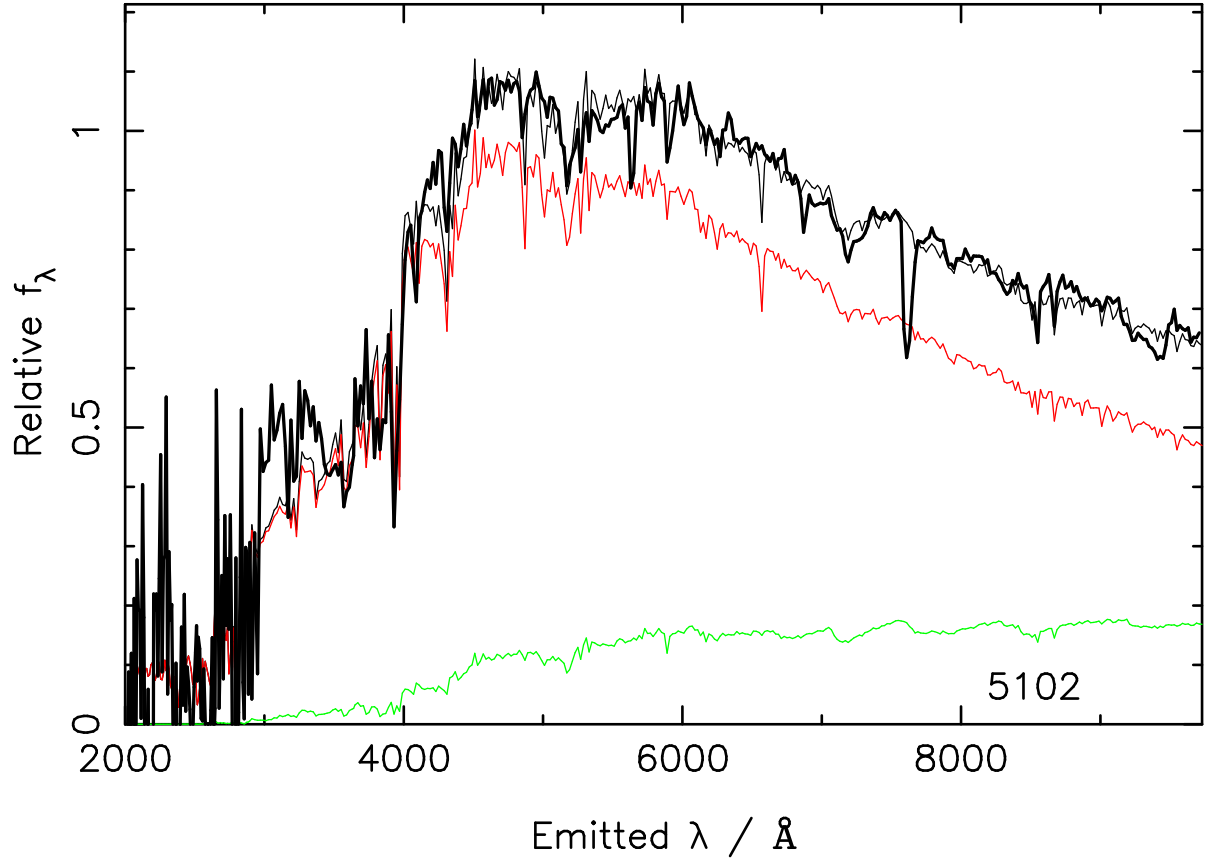}
\caption{The best-fitting two-component model (thin black line) superimposed over the r
esidual of the spectrum of NGC 5102, following subtraction of the 0.3 Gyr population (t
hick black line). The two component populations are also shown; the dominant population
is in red (2 Gyr, Z = \Zsolar), and the lesser population (10 Gyr, Z = 2.5 \Zsolar) is in green.
A faint population of older (10 Gyr) stars is not rejected
by the fitting statistics.}\label{compmodels}
\end{figure}

The age of the older, dominant population in the two-component fit is remarkably young
(3 Gyr). In order to check the possibility that there is a sub-population, present within
this component, of older (10 Gyr) stars, we have fitted a two-component model to the
residual of the spectrum of NGC 5102, following subtraction of the 0.3 Gyr population.
Figure~\ref{compmodels} shows this residual spectrum, with the best-fitting two-component
model overlaid (as in figure \ref{specfig}). The addition of a third, older 
(10 Gyr, 2.5 \Zsolar) stellar population is not rejected by the fitting statistics, 
although the increase in the complexity of the model means that the age 
and metallicity of the third
component are not strongly constrained, and we are potentially over-fitting the data.
The super-solar abundance found in the oldest component is consistent with
other recent estimates of abundances in lenticular galaxies in poor
environments \citep{kun98,tra00,kun02a} and is therefore not implausible.
In order to match the slope of the observed spectrum, the presence of
the 10 Gyr population forces the best-fitting younger population to be even younger 
(2 Gyr). However, even with the addition of the third, oldest component, a substantial
($>\sim$55\%) component of the stars must still be young, and relatively high metallicity
 (\Zsolar) in order to fit both the over-all shape of the continuum and the detailed absorption
features, especially longwards of 5000 \AA, where contributions from more
mature (i.e. $>$ 1 Gyr) stars dominate the spectrum, for example, the TiO absorption
feature at $\sim$7150 \AA. The high metallicity determined for both these two 
sub-populations indicates the presence of strong absorption features, unambiguously dependent
on the metallicity.  It should be noted that the feature at $\sim$ 7500 \AA~is a sky line,
which is excluded from the fitting process.
Therefore, independent of the model chosen (i.e. either 2-component or 3-component), a large
fraction (55-97\%) of the stars in NGC 5102 are $<$3 Gyrs old.  If the XRB population
is linked directly to the stellar population, this would suggest that NGC 5102 has
not had sufficient time to develop an XRB population appropriate to its
stellar mass.

We estimate the mass of the youngest (0.3 Gyr) population to be $\sim$2.2$\times$10$^{7}$ \Msolar.
Again, this is consistent with independent
estimates in the literature. \citet{pri79}, using UBVR photometry,
estimates that $\sim$2$\times$10$^{7}$ \Msolar\ stars formed in last
$\sim$0.1 Gyr. However, Pritchet's estimate is based on an older
estimate of distance, D$=$4.4 Mpc, from \citet{deV75}.
Using this distance in our more sophisticated mass
estimate, we find $\sim$4.4$\times$10$^{7}$ \Msolar.
An HI survey of NGC 5102 detected a ring of emission with
a central depression \citet{woe93}.
It is interesting that the estimate of the
`missing' mass of this depression is $\sim$1.5$\times$10$^{7}$ \Msolar\
(D$=$3.1 Mpc). Although the radial distributions
of the HI depression and central starburst differ, the correlation in
the masses is suggestive of a link between the HI hole and the
formation of the young stellar population. Alternatively, the
starburst occurring 0.3 Gyr ago could have been triggered by the
capture of a gas-rich dwarf galaxy. There are no sizeable galaxies
that are near neighbors with NGC 5102, and it is unlikely that any
star-formation-inducing interaction with another large galaxy could
have occurred within the last 4 Gyr \citep{woe93}.

\subsection{Globular clusters in NGC~5102}

We analyzed three archival HST/WFPC2 observations
that lie contiguously along the major axis of NGC~5102,
obtained with the F569W filter (PI: Freeman). 
The central pointing consisted of 8 exposures of 500s each,
whereas the other two had three exposures of 1100s each. The
observations had large overlaps. The effective area we could survey
for globular clusters, excluding the central saturated region of the
galaxy, was approximately 9.5 sq. arcmin, in a rectangular shape along
the major axis. 
This covered approximately half the optical
light of the galaxy covered by the $D_{25}$ ellipse, or $\sim$46\% of the
total light of the galaxy.
We combined the exposures for each field using the DRIZZLE algorithm
\citep{drizzle}.  We then smoothed each combined image $I$
by running a median filter of 16 pixels square, and subtracted a
scaled version of the smoothed image $S$ from the original image
according to $\alpha I- (\alpha-1) S$, to produce a unsharp masked
image (after trial and error, we chose $\alpha=16$). Finally, we ran
the image finding algorithm SEXTRACTOR to find images that are
significantly non-stellar compared to the local PSF, and visually
inspected the candidate images.

At the distance of NGC~5102, the mean (King) core radius and tidal
radius of a globular cluster (GC) would be approximately 0.15 and 2.8
arcsec. We compared our candidates with those in \citet{holland99},
who studied 5 HST/WFPC2 images in the central region of the giant
elliptical NGC~5128 (Cen~A), at roughly the same distance (3.4 Mpc) as our galaxy. The
\citet{holland99} study covered 25 sq. arcmin (6\% of the $D_{25}$
ellipse) of NGC~5128, which is 16 times more luminous compared to
NGC~5102. In Cen~A, they found 21 globular cluster candidates
($r<2''$) and 61 other extended objects
($r>2''$), some of which could be single or double GCs.
Applying their criterion to our images, we obtained three GC candidates.
The positions and ellipticities of these three objects are tabulated
in Table~\ref{gctab}.
Scaling by the fraction of galactic
light contained in the HST FOV,
we estimate that there are $\sim$7 GCs around NGC 5102.

The globular cluster specific frequency (GCSF) is known to be a strong function of both
environment and morphological type (see \citet{har91} and
references therein).  In particular, elliptical galaxies generally
have larger GC specific frequencies than lenticular galaxies,
and those in richer environments also tend to have larger specific frequencies as well.  
In addition, the members of the Virgo (e.g. NGC 4472) and Fornax clusters 
are known to have exceptionally large GCSF values, even for elliptical
galaxies in rich environments \citep{ber82}.
NGC 5102 is a lenticular galaxy, and
it resides in a poor environment (it is the fourth most massive galaxy
in the Centaurus A group after Cen A, M83, and NGC 4945).
The GCSF, $S_N$, for NGC 5102 is 0.4 if there are 7 GCs,
atypically low for lenticular galaxies in poor environments \citep{har91,kum93}.
A more typical value for the GCSF of $\sim$1.5 given the luminosity, morphological
type, and environment would imply that there would be $\sim$27 GCs.
If the number of XRBs scales as the number of GCs (i.e. if all XRBs in 
early-type galaxies are created in GCs), the deficit of XRBs
in NGC 5102 can be accounted for by its low GCSF.
For example, there are approximately 1000 GCs around Cen A \citep{har04} and
110 LMXBs within 9$'$ with $L_X>$10$^{37}$ ergs s$^{-1}$, or
0.11 XRBs per GC.  Thus, we would expect 1 XRB in NGC 5102 given 7 GCs.
A similar scaling using XRB and GC populations of
NGC 4472 results in the same conclusion.

\subsection{High Mass X-ray Binaries}

Given the recent star formation in NGC 5102, it is possible that some or
all of the XPSs are actually HMXBs; either wind accretors with main sequence
secondaries like Cen X-3 or
Cyg X-1, or Be star binaries like X Per.  The relatively low
luminosity ($\sim$10$^{36}$ ergs s$^{-1}$) of the majority of XRBs in NGC 5102
is more consistent with the population of HMXBs in our galaxy, than of
the LMXBs \citep{gri02}.
For galaxies with a large on-going star formation rate,
it has been argued that a significant fraction of the XPS
population are HMXBs \citep{gri03}.  
The lifetime of objects such as Cyg X-1 and Cen X-3 is
short (a few$\times$10$^7$ yrs), but not less than the estimated
time since the last burst of star formation in
NGC 5102 ($\sim$1.5$\times$10$^7$ yrs \citep{deh97}).
Such objects are rare in our galaxy in large
part because of their short lifetimes and the fact that
they require the original masses of
both progenitor stars to be fairly large \citep{ver96,ver01}.

It is more likely that if any of the objects in NGC 5102 are HMXBs, they are
Be star X-ray binaries.  Given the transient nature of such sources,
it is possible that there is a large population of such objects
in NGC 5102 that are currently not detectable.
\citet{deh97} resolved a population of $\sim$30 blue stars, mostly likely
B main sequence stars, but possibly post-AGB stars, in the central
7$''$ of the galaxy with the HST Faint Object Camera.
The distribution of the stars in this region
follows the light distribution, so that even beyond this central
region, it is possible that the sources are HMXBs.
A correlation between the XPSs and the HII regions would support
this, but as stated above, there are no such coincidences.
A detailed optical follow-up of the X-ray sources would resolve this issue.
We emphasize that if some or all of the XPSs in NGC 5102 turn out to be
HMXBs, this makes the LMXB discrepancy between NGC 5102 and more massive
ellipticals even larger as it is unlikely that a significant fraction of
the XPSs in this latter group are HMXBs.

High mass X-ray binaries with an OB supergiant companion have been observed
in our galaxy to have large (60-100 km s$^{-1}$) velocities \citep{che97}.
Such an object will only be marginally bound
to this relatively low mass galaxy, 
and would travel $\sim$1 kpc in 10$^7$ yrs.
It is therefore possible that one (or more) of the XPSs in the halo
NGC 5102 are such objects.  Be stars are generally created with smaller
velocities, but may be somewhat longer lived.  These halo sources could be Be star binaries
as well.  We noted above that there is a group of seven sources several kpc to the
north of the nucleus with no obvious counterparts in the J-band 2MASS image (see Figure~\ref{psovl}).
One or more of these may be such a halo HMXB ejected from the galaxy.
Again, a detailed optical follow-up of the sources would confirm or refute this.

\section{Discussion}

We have demonstrated that the previously known low X-ray luminosity of NGC 5102 is
the result of a deficit of XRBs coupled with a modest amount
of hot gas.  We have also shown that more than half
of the stars in this unusually blue lenticular galaxy are less than
2-3 Gyrs old, and that NGC 5102 has an unusually small number
of globular clusters.  Is the lack of XRBs related to the relative
youth of the stellar population, to the lack of GCs, or both?  
It is impossible to draw any definitive conclusion
based on a sample of one galaxy and the small number of X-ray sources detected.
There is little doubt that GCs play an important role in the formation
of XRBs in early-type galaxies.  The question, though, is what role
does the stellar population play.
It is probable that both the star formation history and the deficit
of GCs play a role in the lack of XRBs in NGC 5102.

Even if 30-60\% of the XRBs in all early-type galaxies are contained within GCs,
the other 40-70\% of the population
remains unaccounted for.  It is possible that all XRBs in early-type
galaxies were formed in GCs and have been ejected,
although there is little evidence to either support or refute this claim.
On the other hand, this non-GC
XRB population may represent an entirely distinct group of sources.
Given the LMXB population in the Milky Way and M31 bulges, it
would be surprising if most of these sources were not
formed via the stellar evolution scenario described above.
A study of the XRB populations of a sample of early-type galaxies with different
star formation histories is required to address this question.
If the XRB population is directly linked to the star formation
history, our results would
suggest that the stellar population must evolve sufficiently for a significant
XRB population to form, and that there is a minimum timescale of a few
Gyrs for XRB activity to
commence.  This timescale is roughly consistent with estimates of the timescale
required for rotational braking of the magnetic winds of
companion stars to extract orbital angular momentum and therefore reduce
the orbital separation
of a post-supernova binary to the point where Roche lobe overflow occurs
and an LMXB is formed \citep{ver81}.

One important, more general consequence of a relationship
between the LMXBs and the star formation history
would be that the X-ray luminosity of a galaxy should
peak several gigayears after any massive burst of star formation \citep{whi98}.
Most of the star formation in galaxies is believed to have taken
place at z$\sim$1-2 or earlier \citep{bla99}.  
\citet{whi98} calculate that the X-ray luminosity of galaxies should
peak at z$\sim$0.5-1.  Chandra observations of the Hubble Deep Field North
of the X-ray flux from optically bright spiral galaxies support this hypothesis
\citep{bra01,pta01}, which can be generally extended to galaxies
of all morphological types \citep{gho01}.
This temporal evolution of the XRB population
would be hard to explain if it were entirely linked to the GC population.
Clearly there is an important relationship between the
LMXBs and the GC population for the most massive elliptical galaxies
in rich environments with large GCSFs.
However, the SFR history and binary evolution must also
play an important role in the formation of LMXBs.
We are currently investigating the relationship between the XRB population,
the star formation history (via the spectral synthesis technique),
and the globular cluster population in a large sample
of early-type galaxies, the results of which will be presented in a future
publication.  

\acknowledgements

This work was funded by NASA grant GO2-3109X.

\clearpage

\clearpage

\clearpage

\clearpage

\clearpage

\clearpage

\clearpage

\clearpage

\begin{table}
\begin{center}
\begin{tabular}{|c|c|}\hline
Parameter & Value \\ \hline
\multicolumn{2}{|c|}{Position (J2000)} \\ \hline
RA  & 13$^h$21$^m$57.6$^s$ \\ \hline
Dec & -36$^d$37$^m$49$^s$  \\ \hline
\multicolumn{2}{|c|}{Optical Properties} \\ \hline
$m_B$ &  9.99 \\ \hline
$B-V$ & 0.64 \\ \hline
Distance & 3.1$\pm$0.15 Mpc \\ \hline
$L_B$ &  1.49$\times$10$^9$ L$_\odot$ \\ \hline
\multicolumn{2}{|c|}{Chandra Observation Log} \\ \hline
Date & 21MAY02 \\ \hline
Exposure Time & 34217 s \\ \hline
OBSID & 2949 \\ \hline
\end{tabular}
\caption{Summary of parameters for NGC 5102.}\label{n5102tab}
\end{center}
\end{table}

\clearpage

{\footnotesize
\begin{center}
\begin{longtable}{|r|c|c|c|c|}\hline\hline
 Source & RA & DEC & Rate (10$^{-4}$ cts s$^{-1}$) & log(L$_X$) \\ \hline
   1 & 13:21:36.16 & -36:35:48.8 & 2.9 $\pm$ 1.0 &       \\ \hline
   2 & 13:21:38.97 & -36:34:05.9 & 5.8 $\pm$ 1.5 &       \\ \hline
   3 & 13:21:39.07 & -36:38:40.7 & 4.5 $\pm$ 1.2 &       \\ \hline
   4 & 13:21:39.19 & -36:40:32.9 & 9.3 $\pm$ 2.3 &       \\ \hline
   5 & 13:21:39.91 & -36:39:37.0 & 4.9 $\pm$ 1.3 &       \\ \hline
   6 & 13:21:40.36 & -36:40:04.3 & 2.2 $\pm$ 0.9 & 36.24 \\ \hline
   7 & 13:21:42.99 & -36:37:33.5 & 2.0 $\pm$ 0.8 &       \\ \hline
  8 & 13:21:44.35 & -36:36:34.5 & 14.9 $\pm$ 2.2 &       \\ \hline
   9 & 13:21:44.56 & -36:40:37.1 & 4.2 $\pm$ 1.2 & 36.53 \\ \hline
  10 & 13:21:44.86 & -36:34:53.1 & 1.6 $\pm$ 0.7 &       \\ \hline
 11 & 13:21:46.29 & -36:33:54.1 & 14.9 $\pm$ 2.2 &       \\ \hline
  12 & 13:21:46.38 & -36:40:10.5 & 1.7 $\pm$ 0.7 & 36.15 \\ \hline
  13 & 13:21:46.52 & -36:34:41.7 & 2.4 $\pm$ 0.9 &       \\ \hline
  14 & 13:21:46.67 & -36:35:07.1 & 5.5 $\pm$ 1.3 &       \\ \hline
  15 & 13:21:47.66 & -36:36:15.9 & 1.2 $\pm$ 0.6 &       \\ \hline
 16$^\dag$ & 13:21:47.88 & -36:41:34.6 & 10.2 $\pm$ 2.0 &       \\ \hline
  17 & 13:21:48.76 & -36:39:21.7 & 2.3 $\pm$ 0.9 & 36.28 \\ \hline
  18$^\dag$ & 13:21:51.86 & -36:30:50.8 & 2.3 $\pm$ 1.1 &       \\ \hline
  19 & 13:21:52.12 & -36:33:09.1 & 7.9 $\pm$ 1.7 &       \\ \hline
  20 & 13:21:53.04 & -36:32:47.9 & 7.0 $\pm$ 1.6 &       \\ \hline
  21 & 13:21:53.94 & -36:35:38.5 & 3.1 $\pm$ 1.0 &       \\ \hline
  22$^\dag$ & 13:21:54.98 & -36:41:42.9 & 6.1 $\pm$ 1.5 &       \\ \hline
  23 & 13:21:55.16 & -36:35:06.9 & 9.0 $\pm$ 1.6 &       \\ \hline
  24 & 13:21:55.37 & -36:35:41.1 & 3.7 $\pm$ 1.1 &       \\ \hline
  25 & 13:21:55.43 & -36:35:46.0 & 6.3 $\pm$ 1.4 &       \\ \hline
  26$^\dag$ & 13:21:56.78 & -36:41:28.8 & 6.8 $\pm$ 1.6 &       \\ \hline
 27 & 13:21:57.40 & -36:37:51.0 & 33.8 $\pm$ 3.2 & 37.44 \\ \hline
  28 & 13:21:57.63 & -36:37:48.8 & 4.4 $\pm$ 1.2 & 36.55 \\ \hline
  29 & 13:21:57.67 & -36:38:36.5 & 1.1 $\pm$ 0.6 & 35.97 \\ \hline
  30 & 13:21:57.96 & -36:34:57.3 & 3.9 $\pm$ 1.1 &       \\ \hline
  31$^\dag$ & 13:21:59.30 & -36:41:08.8 & 1.9 $\pm$ 0.9 &       \\ \hline
  32 & 13:21:59.33 & -36:38:20.3 & 1.7 $\pm$ 0.7 & 36.13 \\ \hline
  33 & 13:21:59.37 & -36:34:19.5 & 2.1 $\pm$ 0.8 &       \\ \hline
  34 & 13:21:59.76 & -36:34:46.0 & 5.6 $\pm$ 1.3 &       \\ \hline
  35$^\dag$ & 13:22:00.35 & -36:41:33.7 & 3.4 $\pm$ 1.2 &       \\ \hline
  36 & 13:22:01.05 & -36:39:00.1 & 1.7 $\pm$ 0.7 & 36.15 \\ \hline
  37$^\dag$ & 13:22:01.59 & -36:30:38.4 & 9.0 $\pm$ 2.4 &       \\ \hline
  38$^\dag$ & 13:22:04.45 & -36:42:20.0 & 1.5 $\pm$ 0.8 &       \\ \hline
  39 & 13:22:04.91 & -36:38:54.6 & 5.2 $\pm$ 1.2 &       \\ \hline
  40$^\dag$ & 13:22:06.05 & -36:30:49.1 & 108.4 $\pm$ 11.0 &       \\ \hline
  41 & 13:22:06.62 & -36:34:39.1 & 1.4 $\pm$ 0.7 &       \\ \hline
  42$^\dag$ & 13:22:06.89 & -36:40:02.0 & 3.6 $\pm$ 1.2 &       \\ \hline
  43 & 13:22:07.32 & -36:34:50.9 & 5.9 $\pm$ 1.4 &       \\ \hline
  44 & 13:22:07.32 & -36:33:23.7 & 2.3 $\pm$ 0.9 &       \\ \hline
  45 & 13:22:07.68 & -36:34:00.1 & 1.9 $\pm$ 0.8 &       \\ \hline
 46 & 13:22:10.94 & -36:36:49.0 & 12.4 $\pm$ 2.0 & 37.00 \\ \hline
  47$\dag$ & 13:22:11.71 & -36:40:34.0 & 1.9 $\pm$ 0.9 &       \\ \hline
  48$\dag$ & 13:22:12.45 & -36:40:10.7 & 1.1 $\pm$ 0.7 &       \\ \hline
  49 & 13:22:12.71 & -36:38:02.8 & 7.5 $\pm$ 1.5 &       \\ \hline
 50 & 13:22:12.76 & -36:37:52.5 & 11.6 $\pm$ 1.9 &       \\ \hline
  51 & 13:22:14.50 & -36:35:28.1 & 3.6 $\pm$ 1.3 & 36.46 \\ \hline
  52$^\dag$ & 13:22:14.97 & -36:42:43.5 & 8.2 $\pm$ 2.0 &       \\ \hline
 53$^\dag$ & 13:22:15.49 & -36:41:57.2 & 27.7 $\pm$ 3.5 &       \\ \hline
  54 & 13:22:16.09 & -36:39:20.6 & 1.9 $\pm$ 1.1 &       \\ \hline
  55$^\dag$ & 13:22:16.40 & -36:41:03.0 & 3.2 $\pm$ 1.2 &       \\ \hline
\caption{Summary of X-ray point sources around NGC 5102.
The count rates and uncertainties are in the 0.5-5.0 keV band.  The X-ray luminosities
are given for those sources that are within the optical $D_{25}$ isophote (see text for
complete discussion) assuming a distance of 3.1 Mpc.  The sources labeled with
a dagger lie on the S2 or S4 chips, none of which are within the
$D_{25}$ isophote.}\label{bigtab}
\end{longtable}
\end{center}
}

\clearpage

\begin{table}
\begin{center}
\begin{tabular}{|c|c|c|c|c|c|c|c|c|}\hline
Galaxy & $m_B$ & $m_J$ & Distance (Mpc) & $M_B$ & $M_J$ & $N_{XPS}$ \\ \hline
NGC 4472 & 9.33 & 6.27 & 16.0 & -21.69 & -24.75 & $\sim$220 \\ \hline 
NGC 5128 & 7.30 & 4.98 &  3.5 & -20.24 & -22.76 & $\sim$110 \\ \hline\hline
NGC 5102 & 9.97 & 7.71 &  3.1 & -17.49 & -19.75 & 5(est.)/1(obs.) \\ \hline
\end{tabular}
\caption{Comparison of optical/IR properties of NGC 5102, NGC 5128 (Cen A), and
NGC 4472.}\label{galsum}
\end{center}
\end{table}

\clearpage

\begin{table}
\begin{center}
\begin{tabular}{lcl}
\hline
{source} & {wavelength range - \AA}  \\
\hline
IUE Newly Extracted Spectra & 2000-3200   \\
 Bica \& Alloin, unpublished     &      3100-5400  \\
 \citet{bic87a} & 3800-7500  \\
 \citet{bic87b} & 6400-9800  \\
\hline
\end{tabular}
\caption{Sources of the archival spectra of NGC 5102. The Bica \& Alloin data are
available at ftp://cdsarc.u-strasbg.fr/cats/III/219/.}\label{datatable}
\end{center}
\end{table}

\clearpage

\begin{table}
\begin{center}
\begin{tabular}{|c|c|c|c|c|}\hline\hline
 {age / Gyr} & { Z / \Zsolar } & {M / M$_{gal}$} & {minimum \xsnu} & {$k$}\\
\hline
 3.0  &  1.5 & 0.93  &  174.7/430  & 0.10 \\
 0.3  &  0.2 & 0.07  & &  \\
\hline
\end{tabular}
\caption{The results of fitting the near-UV-to-optical spectra of NGC 5102 with
 the two-component model spectrum. The parameters quoted are those values on the
age-metallicity grid which correspond to the minimum calculated \xs. The age
and metallicity of the two-component populations are given, for the best fitting
combination for each model set. M / M$_{gal}$, the fractional contribution, by
mass, to the total population made by each component is shown in column 3, and
the dust extinction parameter, $k$, is given in column 5.}\label{spectable}
\end{center}
\end{table}

\clearpage

\begin{table}
\begin{center}
\begin{tabular}{|l|c|c|}\hline\hline
\multicolumn{3}{|c|}{Globular cluster candidates} \\ \hline
13:21:51.73 & -36:38:56.9 & 0.066 \\ \hline
13:21:47.33 & -36:38:11.2 & 0.149 \\ \hline
13:21:56.71 & -36:37:34.1 & 0.261 \\ \hline
\end{tabular}
\caption{Positions (J2000) and ellipticities ($1-b/a$) of globular cluster
candidates detected in HST observations.}\label{gctab}
\end{center}
\end{table}

\end{document}